\documentclass[a4,useAMS,usenatbib,usegraphicx]{mn2e}
\def\apss{\ref@jnl{Ap\&SS}} 
\def\aapr{\ref@jnl{A\&A~Rev.}}
\def\na{\ref@jnl{New~Astronomy}}

\usepackage{amsmath}
\include{epsf}
\usepackage{float}
\usepackage{color}
\usepackage{multirow}

\usepackage{epsf}
\epsfclipon

\title[Ram Pressure on Tidal Dwarf Galaxies]{The Influence of Ram Pressure on the Evolution of Tidal Dwarf Galaxies}
\author[R. Smith et~al.]{R. Smith$^{1}$\thanks{E-mail:rsmith@astro-udec.cl}, P. A. Duc${^2}$, G. N. Candlish${^1}$, M. Fellhauer${^1}$, Y. K. Sheen$^{1}$, B. K. Gibson${^3}$ \\
$^{1}$Departamento de Astronomia, Universidad de Concepcion, Casilla 160-C, Concepcion, Chile\\
$^{2}$Laboratoire AIM, Service d'astrophysique, CEA Saclay, Orme de Merisiers, Batiment 709, 91191 Gif sur Yvette cedex, France\\
$^{3}$Jeremiah Horrocks Institute, University of Central Lancashire, Preston, PR1~2HE, United Kingdom}
\begin{document}

\date{\today}

\pagerange{\pageref{firstpage}--\pageref{lastpage}} \pubyear{2013}

\maketitle

\label{firstpage}

\begin{abstract}
The formation mechanism of tidal dwarf galaxies means they are expected 
to contain little or no dark matter. As such, they might be expected to 
be very sensitive to their environment. We investigate the impact of ram 
pressure on tidal dwarf galaxies in a parameter study, varying dwarf 
galaxy properties and ram pressures. We submit model tidal dwarf 
galaxies to wind-tunnel style tests using a toy ram pressure model. The 
effects of ram pressure are found to be substantial. If tidal dwarf 
galaxies have their gas stripped, they may be completely destroyed. Ram 
pressure drag causes acceleration of our dwarf galaxy models, and this further 
enhances stellar losses. The dragging can also cause stars to lie in a 
low surface brightness stellar stream that points in the opposite 
direction to the stripped gas, in a manner distinctive from tidal 
streams. We investigate the effects of ram pressure on surface density 
profiles, the dynamics of the stars, and discuss the consequences for 
dynamical mass measurements.
\end{abstract}


\section{Introduction}
Tidal dwarf galaxies (TDGs) are, by definition, bound structures formed 
from tidal tails of gas and stars ({\citealp{Duc2012}}). The tidal tails 
are produced by tides or gravitational torques when progenitor galaxies 
merge or gravitationally interact. In the nearby Universe, the 
progenitors are typically two late-type galaxies, whereas TDG formation 
about early type galaxies is very rare (\citealp{Kaviraj2012}). There 
are at least two mechanisms that can form substructures within tidal 
tails with masses equal to dwarf galaxies 
($\sim$10$^{7}$$-$10$^{9}$~M$_\odot$). The first is collapse through 
Jeans instabilities -- when an object has sufficient mass to overcome 
internal support provided by pressure, and collapses to form a 
self-bound object (\citealp{Barnes1992}). \citet{Wetzstein2007} find 
that the presence of gas is required for TDGs to form in this manner. 
The result is numerous star-forming gas clumps distributed along a tidal 
tail. The second is where a large region of the outer disk material 
tends to pool at the end of a tidal tail, and becomes self-bound 
(\citealp{Elmegreen1993}). This second formation mechanism tends to 
produce a single, massive, TDG. Simulations have demonstrated that this 
type of TDG forms more readily when the progenitor galaxies have large, 
extended, dark matter halos. If they did not, numerous, low mass, 
`beads-on-a-string' tend to form (\citealp{Bournaud2003}; 
\citealp{Duc2004}). In fact, observed TDGs appear to either form large 
numbers of low mass galaxies, or a single massive object 
\citep{Knierman2003}.

While TDGs remain associated with their natal tidal tails, they are 
easily identified. However with sufficient time, some TDGs may no longer be seen in close proximity to tidal features. Such TDGs may be 
difficult to differentiate from more `typical' dwarf irregulars (dIrrs), 
as they have similar luminosities, morphologies, gas fractions, and 
sizes (\citealp{Hunter1997}; \citealp{Duc2012}). A number of possible 
approaches to identify long-lived TDGs are compared and discussed in 
\citet{Hunter2000} and \citet{Duc2012}. As TDGs are formed from the 
recycled gas of more massive (and metal-rich) progenitors, they are 
expected to show elevated metallicities compared to other star forming 
dwarfs of similar luminosity. Even when TDGs form in the very outskirts 
of the progenitor galaxies, their metallicity remains high 
\citep{Weilbacher2003}, perhaps due to strong radial mixing during 
mergers \citep{DiMatteo2011}. A comparison of the dispersion, 
skewness, and kurtosis of the metallicity distribution functions 
of putative TDGs, as opposed to dark matter-dominated dwarfs, may prove 
invaluable as a tool to discriminate between these competing scenarios 
\footnote{For example, dwarf galaxy IKN in the M81 group, a potential TDG 
candidate \citep{Lianou2010}, possesses a negative skewness somewhat in 
excess of the other dwarf spheroidals in the group; whether this is 
indicative of a fundamental difference, or simply a coinicidence, should 
be assessed, using the tools of, say, \citet{Pilkington2012b}.}. Another 
clue may come from the spectral energy distribution of star forming 
regions in TDGs, as they may be forming their first generation of stars 
but, at the same time, be chemically evolved \citep{Boquien2010}.

A clear distinction between typical dIrrs and TDGs lies in their dark 
matter content. TDGs should contain little or no dark matter 
(\citealp{Bournaud2010}). This may be detectable in the behaviour of 
their rotation curves, or in terms of a deviation from the Tully-Fisher 
relationship of normal, dark matter-dominated, dwarfs 
(\citealp{Hunter2000}). For a few TDGs, the mass budget has been 
measured in terms of the total stellar, atomic, and molecular gas 
detected; in those cases, a small amount of additional `missing' mass, 
not accounted for in this detected mass budget, is required in order to 
explain the internal dynamics, under the assumption of dynamical 
equilibrium. Explanations include the presence of `dark gas' 
(\citealp{Bournaud2010}) or modified Newtonian dynamics 
(\citealp{Boquien2010}).

The lack of a protective dark matter halo could cause TDGs to be 
extremely sensitive to their environment. In the absence of resonant 
stripping (\citealp{Muccione2004}; \citealp{Donghia2009}), dwarf 
galaxies which initially have massive dark matter halos are protected 
from stellar mass loss due to external tides, until $\sim$80-90$\%$ of 
their dark matter has been stripped (\citealp{Knebe2006}; 
\citealp{Penarrubia2008}; \citealp{Smith2013}). Therefore, TDGs which 
have no dark matter from the beginning, should be susceptible to tidally 
induced mass loss. Damaging tides could initially arise from the gravity 
of the progenitor galaxies. Simulations suggest a fraction of the 
initially formed TDGs can survive tidal destruction from their 
progenitors in order to become long-lived TDGs (\citealp{Bournaud2003}; 
\citealp{Bournaud2010}). TDGs that survive this first hurdle and, 
additionally, avoid infalling back onto their progenitors, must face the 
tides of the group or cluster environment in which they find themselves.

Besides tides, TDGs may also be susceptible to ram pressure stripping. 
The effects of ram pressure on the behaviour of the stars of a galaxy is 
generally assumed to be very minor in giant late-type galaxies. This is 
because the cross-sectional area of stars and molecular clouds is too 
small to feel any significant acceleration directly from the hot gas 
that causes ram pressure. However, the removal of this gas can lower the 
overall disk mass. This may result in some thickening of the stellar 
disk (\citealp{Farouki1980}). However, this is generally assumed to be 
minor as the gas fraction is typically low ($\sim$10\%) in nearby giant 
late-type disk galaxies (\citealp{Gavazzi2008}).

At lower masses, however, late-type dwarf galaxies can contain very high 
gas fractions, with several times more gas than stars 
(\citealp{Gavazzi2008}). In such galaxies, the removal of the gas mass 
changes significantly the galaxy's potential, causing thickening of the 
stellar disk by roughly a factor of two (\citealp{Smith2012}).

In fact, the effect of ram pressure on the stars goes beyond disk 
thickening. In \cite{Schulz2001}, it was noted that the stellar disc of 
their models was displaced several kiloparsecs downwind, in their ram 
pressure simulations. This effect was further studied by 
\cite{Smith2012} using simulated late-type dwarf galaxies embedded in 
dark matter halos. Although the stars alone cannot feel directly the ram 
pressure, the gas disc presents a cross-sectional area $A$ to the ram 
pressure wind. As such, it feels a drag force ($F_{\rm{drag}}=A \times 
P_{\rm{ram}}$ where $P_{\rm{ram}}$ is the ram pressure) and is thus 
accelerated in the direction of the ram pressure wind. Meanwhile, the 
gravitational attraction between the stars and the gas disk causes the 
drag force felt by the gas disk to be shared with the stellar disk. In 
fact, the effect is not limited to the stars - the central dark matter 
surrounding the gas disk also feels the drag force. The net result is 
displacement of the gas and stellar disk, and central dark matter, 
downwind of the ram pressure by several kiloparsecs with respect to the 
outer halo. The process is referred to as {\it{ram pressure drag}}. 
\citet{Smith2012} also noted that, because the stellar disk is 
gravitationally dragged by its centre, a short-lived ($<$200~Myr) 
conical-like distortion of the disk results. This is because the stellar 
disk centre is displaced preferentially with respect to the outer 
stellar disk.

The effects of ram pressure drag on TDGs are expected to be even more 
substantial. Like normal late-type dwarfs, TDGs appear to have very high 
gas fractions (\citealp{Duc2000}; \citealp{Hibbard2001}; 
\citealp{Bournaud2004}; \citealp{Mundell2004}; \citealp{Neff2005}; 
\citealp{Bournaud2007}; \citealp{Koribalski2009}; 
\citealp{Bournaud2010}). However, unlike normal late-type dwarfs, TDGs 
contain little or no dark matter. Therefore the loss of the gas 
potential when the gas is stripped should result in stronger disk 
thickening in TDGs than was seen in our dIrr models 
(\citealp{Smith2012}). Also, in normal late-type dwarfs, the ram 
pressure drag force must tow the mass of the stars and dark matter, 
whereas in a TDG it need only tow the mass of the stars. Therefore, for 
the same momentum transfer from the ram pressure wind, a TDG should have 
a greater acceleration.

If TDGs are indeed as sensitive to their environment as proposed, then 
we might expect to see some indication of this in their observed 
properties. \cite{Kaviraj2012} note that the majority of their TDGs are 
located in the field environment. \cite{Sheen2009} detect TDGs in the 
Coma cluster, but specifically in the cluster outskirts. This perhaps 
suggests that the cluster environment is a harsh environment for TDGs, 
but it could also suggest that the high interaction velocities of 
cluster galaxies are not conducive in creating the tidal tails from 
which TDGs form. If the latter, then TDGs may form outside the cluster 
environment, and then be subsequently accreted into the cluster. 
Possible examples include ARP~105 (\citealp{Duc1994}), NGC~5291 
(\citealp{Duc1998b}), IC~1182 (\citealp{Iglesias2003}), and VCC~2062 
(\citealp{Duc2007}). However a dense environment where galaxy relative 
velocities are lower than in clusters (i.e., compact groups) may be 
highly conducive to TDG formation. \cite{Temporin2003} find the compact 
group CG~J1720-67.8 is rich with TDG candidates. From a sample of 42 
Hickson compact groups, \cite{Hunsberger1996} state that perhaps as many 
as half of all the dwarfs in their compact groups are TDGs.

Numerical simulations have provided a powerful tool in studying the 
formation mechanisms of TDGs (e.g., \citealp{Barnes1992}; 
\citealp{Elmegreen1993}; \citealp{Bournaud2003}; \citealp{Duc2004}; 
\citealp{Wetzstein2007}). They have also been used to 
model dynamics in individual, well-studied, interacting systems 
(\citealp{Duc2000}; \citealp{Struck2005}; \citealp{Hancock2007}). 
\cite{Recchi2007} use 2D chemo-dynamical models to understand how TDGs 
react to the feedback from ongoing star formation, and find that star 
formation can continue for more than 300~Myr. In high-resolution cluster 
formation simulations, \cite{Puchwein2010} finds that up to $\sim$30$\%$ 
of their intracluster stars are formed extra-galactically in stripped 
gas clouds. With increasing resolution, it has recently become possible 
to better study the properties of TDGs formed in galaxy 
interaction/merger simulations such as mass, internal dynamics, or even 
rotation curves (\citealp{Bournaud2008I}).

Here we use numerical simulations to study the response of model TDG 
galaxies to gas removal by ram pressure stripping, and ram pressure 
drag. Even if TDGs form outside the cluster, they may later be accreted 
into the cluster. These TDGs will then be subjected to ram pressure from 
their motion through the intracluster medium. However ram pressure is 
not limited to occurring in the cluster environment. Although the 
densities of the hot gas appear lower (\citealp{Mushotzky2004}), dwarf 
galaxies moving through the intra-{\it{group}} medium may also suffer 
ram pressure (\citealp{Marcolini2003}). In fact, even individual 
galaxies contain their own hot gaseous halos that may exert ram pressure 
on other galaxies that move through them. A hot gaseous halo is seen 
surrounding the Milky Way (\citealp{Bregman2007}; \citealp{Lehner2011}; 
\citealp{Gupta2012}), and also in other galaxies 
(\citealp{Tumlinson2011}; \citealp{Tripp2011}). Ram pressure from the 
Milky Way's (MW's) hot gaseous halo has been used to explain the 
morphology-density relation of dwarf galaxies surrounding the MW 
(\citealp{Mayer2005}). Recently the argument was reversed in 
\cite{Gatto2013}. By assuming that ram pressure caused the observed halt 
in star formation in the dwarf spheroidals Sextans and Carina, limits 
were placed on the density of the hot gaseous halo of the MW. That 
derived density is in broad agreement with previous measurements. The 
presence of the MW's hot halo is also seen indirectly through its action 
on the Magellenic stream (\citealp{Weiner1996}; 
\citealp{Stanimirovic2002}; \citealp{Putman2003a}). The existence of hot 
gaseous halos is expected in the galaxy formation framework of 
$\Lambda$CDM (\citealp{Feldmann2013}), and may be an important 
ingredient in forming disks with more realistic morphologies 
(\citealp{Hambleton2011}; \citealp{Brook2012}).

To study the effects of ram pressure on TDGs, we subject TDG models to 
idealised wind-tunnel style ram pressure tests. The ram pressures and 
properties of the TDG models are varied in a parameter study. We show 
that the effects of ram pressure can be very significant for TDGs -- in 
fact when their gas is stripped, they are completely destroyed. This may have some impact on the survival time of TDGs, a key ingredient in discussions regarding the fraction of dwarf galaxies of tidal origin. We 
describe the set up of our model galaxies, the parameter study, the 
numerical code, and the ram pressure model in \S2. Our results are 
presented in \S3 including baryonic mass loss, drag, stellar streams, 
effects on surface density profiles and stellar dynamics, and 
implications for dynamical mass measurements in TDGs. We discuss our 
results in \S4 and summarise the conclusions in \S5.

\section{Setup}
\subsection{Disk galaxy models}
Our tidal dwarf galaxy (TDG) models consist of two components: 
exponential discs of gas and stars. The methods used to form each 
component will be discussed briefly in the following sections. The true spatial distribution of gas and stars in TDGs is not well constrained observationally, as they are typically poorly resolved. However a disk-like distribution is a reasonable assumption, given the rotating gas disks of TDGs that are both observed (\citealp{Bournaud2007II}), and produced in simulations (\citealp{Bournaud2008I}).

Radially, the stellar and gas disk both have an exponential form
\begin{equation}
\label{expdisk}
\Sigma(R) = \Sigma_0 {\rm{exp}} (-R/R_{\rm{d}})
\end{equation}

\noindent
where $\Sigma$ is the surface density, $\Sigma_{\rm{0}}$ is central 
surface density, $R$ is radius within the disk, and $R_{\rm{d}}$ is the 
scale-length of the disk. For simplicity, we assume the scale-length of 
the stellar disk and gas disk are equal.

We distribute the particles vertically out of the disk using the form 
given by Spitzer's isothermal sheet:
\begin{equation}
\rho(R,z) = \frac{\Sigma(R)}{2z_{\rm{d}}} {\rm sech}\,^2(z/z_{\rm{d}}){\rm{.}}
\end{equation}

\noindent 
Following \cite{SpringelWhite1999}, we make z$_{\rm{d}}$ a fixed fraction of the disk scalelength, and we choose z$_{\rm{d}}$=0.1~R$_{\rm{d}}$.

Disk particles are initially set-up on circular orbits. The circular 
velocity of an exponential disk can be calculated from 
(\citealp{Freeman1970}):
\begin{equation}
v^2_{\rm{c}}(R)=R\frac{d\phi}{dR}=4 \pi G \Sigma_0 R_{\rm{d}} y^2 [I_0(y) K_0(y)-I_1(y) K_1(y)]
\end{equation}
\noindent
where $y$$\equiv$$R$/2$R$$_{\rm{d}}$, and $I$ and $K$ are modified 
Bessel functions. In practice, the term in square brackets is evaluated 
using a look-up table.

The Toomre stability criterion is defined as:
\begin{equation}
\label{Toomreeqn}
Q \equiv \frac{\kappa\sigma_{R}}{3.36 G \Sigma} > 1
\end{equation}

\noindent
where $\Sigma$ is the surface density, $\sigma_{\rm{R}}$ is the radial 
velocity dispersion, and $\kappa$ is the epicyclic frequency defined, 
using the epicyclic approximation (\citealp{SpringelWhite1999}). The azimuthal velocity dispersion $\sigma_{\rm{\phi}}$, and the velocity dispersion out of the plane of the disk $\sigma_{\rm{z}}$, are functions of $\sigma_{\rm{R}}$. We use $\sigma_{\rm{\phi}}^2 = \frac{\sigma_{\rm{R}}^2}{\gamma^2}$ where
\begin{equation}
\gamma^2 \equiv \frac{4}{\kappa^2 R} \frac{d\Phi}{dR} {\rm{,}}
\end{equation}

\noindent
and $\phi_{\rm{z}} = 0.6 \cdot \phi_{\rm{R}}$ \citep{Shlosman1993}. In 
practice, a radially varying velocity dispersion is chosen that 
satisfies $Q>$1.5 at all radii. This ensures that disks are sufficiently 
stable such that their properties do not change significantly for models 
evolved in isolation. This velocity dispersion is physically added to 
the velocities of the star particles. Instead, for gas particles, a 
thermal energy $u$ of the particles is chosen that is equivalent to the 
required velocity dispersion $\sigma$ at that radius 
($u=\sigma^2/(\gamma-1)$ where $\gamma=5/3$ for a monatonic gas). In a 
final step, the rotation velocity $v_\phi$ of all particles in the disk 
is adjusted, accounting for the outwards pressure term caused by the 
gradient in the velocity dispersion:
\begin{equation}
v^2_\phi(R)=v^2_{\rm{circ}}-\frac{R}{\rho}\left|\frac{dP}{dR}\right|
\label{vphipresscorr}
\end{equation}

\noindent where $v_{\rm{circ}}$ is the circular velocity, and other 
terms have their usual meaning. We note that equation 
\ref{vphipresscorr} is strictly only valid in the plane of the disk. 
However, in practice, we find this simplification has negligible 
consequences. We evolve all of our disk models in isolation for at least 
2.5~Gyr to ensure they are dynamically stable, and that any transient 
effects have settled. Therefore, none of our models are placed in 
wind-tunnel tests until they are in dynamical equilibrium. Each TDG 
model is composed of 1$\times$10$^5$ disk particles, split equally 
between gas and stars.

\begin{table*}
\centering
\begin{tabular}{|c|c|c|c|c|c|}
Galaxy model & Total mass & Gas fraction & Effective radius & Central surface density & Wind speeds$^{\star}$ \\
& (M$_\odot$) & & (kpc) & (M$_\odot$~pc$^{-2}$) & (km s$^{-1}$) \\
\hline \hline
M8gf50R1 & 10$^8$ & 0.5 & 1.0 & 15.9 & 200,400,800 \\
M8gf70R1 & 10$^8$ & 0.7 & 1.0 & 15.9 & 200,400,800 \\
M8gf90R1 & 10$^8$ & 0.9 & 1.0 & 15.9 & 200,400,800 \\
\hline
M8gf50R2 & 10$^8$ & 0.5 & 2.0 & 4.0 & 200,400,600 \\
M8gf70R2 & 10$^8$ & 0.7 & 2.0 & 4.0 & 200,400,600 \\
M8gf90R2 & 10$^8$ & 0.9 & 2.0 & 4.0 & 200,400,600 \\
\hline
M8gf50R3 & 10$^8$ & 0.5 & 3.0 & 1.8 & 100,200,300 \\
\hline
M7gf70R1 & 10$^7$ & 0.7 & 1.0 & 1.6 & 100,200,300,400 \\
\end{tabular}
\caption{Main properties of the model tidal dwarf 
galaxies (columns 1-5) and wind-tunnel tests conducted upon them (column 
6). ($^{\star}$In the frame of reference of the TDG, ram pressure occurs 
from incoming wind of hot gas. The quoted velocities are for a ram 
pressure wind with densities typical of the outer Virgo cluster, loose 
galaxy groups, or the inner Milky Way halo.)}
\label{windtest}
\end{table*}

We emphasise that we are not forming TDGs in our simulations -- instead 
we choose to start our wind-tunnel tests with dynamically stable models. 
In reality, it is uncertain if real TDGs are so close to dynamical 
equilibrium. However we wish to examine the effects of ram pressure, 
alone, on the dynamics of TDGs, and it is easier to detect it on models 
that would otherwise be in equilibrium in the absence of ram pressure.

\subsection{Tidal dwarf galaxy models -- a parameter study}

Our parameter study involves varying the mass, disk scalelength, and gas 
fraction of the TDG models. We consider two masses of TDG model: a lower 
mass model and a higher mass variant with a total mass of 
1$\times$10$^7$~M$_\odot$ and 1$\times$10$^8$~M$_\odot$, respectively. 
We consider three disk sizes with effective radii of 1, 2, or 3~kpc, typical of that observed in both young and evolved TDGs (\citealp{Paudel2013}; in prep). As 
TDGs are observed to be typically gas-rich, we consider three gas 
fractions: 50$\%$ (equal gas and stars), 70$\%$, and 90$\%$ (very 
gas-rich).

We subject each model TDG to wind tunnel tests with a fixed wind speed. 
However, we vary the wind speed between the tests, in order to quantify 
a TDG model's response to differing ram pressures. We consider a 
sufficient range of ram pressure to ensure that each model is entirely 
stripped of its gas when subjected to the uppermost wind speed.

When describing a specific model, we use a shorthand label. For example 
M8gf50R1 is the model with total mass of 10$^8$~M$_\odot$ (M8), with a 
gas fraction (gf50) of 50$\%$, and with an effective radius (R1) of 
1~kpc. A list of the key properties of each model TDG, and the wind 
speeds to which they were subjected, can be found in 
Table~\ref{windtest}.

\subsection{The Code}

In this study we use {\sc{gf}} 
(\citealp{Williams2001};\citealp{Williams1998}), a gravitational Tree 
N-body + SPH code that operates primarily using the techniques described 
in \cite{Hernquist1989}. While the Tree code allows for rapid 
calculation of gravitational accelerations, the SPH code allows us to 
include an HI gas component to our tidal dwarf galaxy models. In all 
simulations, the gravitational softening length, $\epsilon$, is fixed 
for all particles at a value of 100~pc. Gravitational accelerations are 
evaluated to quadrupole order, using an opening angle 
$\theta_{\rm{c}}$=0.7. A second-order individual particle time-step 
scheme was utilised to improve efficiency following the methodology of 
\cite{Hernquist1989}. Each particle was assigned a timestep that is a 
power of two division of the simulation block timestep, with a minimum 
timestep of $\sim$5~yr. Assignment of timesteps for collisionless 
particles is controlled by the criteria of \cite{Katz1991}, whereas SPH 
particle timesteps are assigned using the minimum of the gravitational 
timestep and the SPH Courant conditions with a Courant constant $C$=0.1 
(\citealp{Hernquist1989}).

As discussed in \cite{Williams2004}, the kernel radius $h$ of each SPH 
particle was allowed to vary such that at all times it maintains between 
30 and 40 neighbours within 2$h$. In order to realistically simulate 
shocks within the SPH model, the artificial viscosity prescription of 
\cite{Gingold1983} is used with viscosity parameters $(\alpha,\beta)$ = 
(1,2). The equation of state for the gas component of the galaxies is 
adiabatic. We choose a velocity dispersion that varies radially within 
the disk, such that the Toomre stability criteria (Eqn. \ref{Toomreeqn}) 
is satisfied at all radii. For example, for Model M8gf70R1 this requires 
a velocity dispersion of 12 km\,s$^{-1}$ in the disk centre, falling to 
5~km~s$^{-1}$ in the outer disk. This is in good agreement with the 
observed radial variation in HI velocity dispersion in late-type disks 
(\citealp{Tamburro2009}; \citealp{Pilkington2011}). We do not include 
star formation in these models, although we note that stellar 
feedback may further enhance gas stripping by ram pressure (\citealp{Gatto2013}). Our choice 
of an adiabatic equation of state aids us in setting up stable disks. 
Thus, our models are dynamically stable before they are subjected to the 
effects of ram pressure. In this way, we can clearly identify the 
effects of ram pressure on galaxy stability when we subject them to a 
wind tunnel test. Although our simplifying choice of an adiabatic 
equation of state is a gross simplification of a real galaxy's ISM, all 
of our key results occur primarily because of the removal of the gas 
mass by ram pressure. With this in mind, we do not expect that any of 
our key results would change significantly if we had instead used a more 
complex (and numerically expensive) treatment for the ISM, as long as 
the mass of the ISM that is removed by ram pressure is comparable. 
\citet{Tonnesen2009} conduct high resolution simulations of disk 
galaxies undergoing ram pressure in which a multiphase ISM is resolved. 
To first order, the mass of gas stripped from their models does not 
depend sensitively on their ability to resolve a complex multiphase ISM.

\subsection{The Ram Pressure Model}
\label{rpsmodel}

The ram pressure stripping model is very similar to that presented in 
\citet{Vollmer2001} and is the same as the model employed in 
\citet{Smith2012}. In this model, additional acceleration vectors are 
added to individual gas particles to mimic the ram pressure. A live 
intra-cluster medium (ICM) component is not included. For an individual 
gas cloud, moving through the ICM of density $\rho_{ICM}$, with a 
velocity $v$, the pressure on its surface due to sweeping through the 
medium is assumed to be
\begin{equation}
\label{RPSpress}
P_{\rm{ram}} = \rho_{\rm{ICM}} v^2
\end{equation}

\noindent
In order to calculate the strength of the acceleration that gas clouds will 
feel as a result of ram pressure, we follow \cite{Vollmer2001}. A 
constant column density is assumed for each individual cloud. This has 
the advantage that the acceleration due to ram pressure is the same for 
all clouds, disregarding their masses. A value of $\Sigma_{\rm{cld}} = 
7.5 \times 10^{20}$ cm$^{-2}$ ($\sim$6.0 M$_\odot$~pc$^{-2}$) is used. This is comparable with 
measurements made by \cite{Rots1990} and \cite{Crosthwaite2000} on 
nearby face-on galaxies. The acceleration due to ram pressure can 
therefore be written as
\begin{equation}
\label{RPSforce}
a_{\rm{ram}} = \frac{P_{\rm{ram}}}{m_{\rm{H}}\Sigma_{\rm{cld}}}
\end{equation}

\noindent
where $m_{\rm{H}}$ is the mass of a hydrogen atom. In the galaxy model's 
frame of reference, the effect of its motion through the ICM is that of 
an ICM wind of velocity $v$ and density $\rho_{\rm{ICM}}$; $a_{\rm{ram}}$ 
therefore always acts in the direction of the velocity vector of the 
wind.

Once more, following \cite{Vollmer2001}, a shading criteria is used to 
select gas particles that feel the influence of ram pressure, and gas 
particles that are shielded by other gas particles upstream in the wind. 
In the simulation, gas particles are point particles and therefore have 
no cross-section by which to shield other particles. However, we can 
calculate a particle's cloud radius $r_{\rm{cld}}$ if we know its mass 
$m_{\rm{p}}$ and once more assume it has the column density 
$\Sigma_{\rm{cld}}$. Then $r_{\rm{cld}}$= $\left[m_{\rm{p}}/(\pi 
m_{\rm{H}} \Sigma_{\rm{cld}})\right]^{0.5}$. In practice, each particle 
extends an imaginary vector along the direction of motion of the 
galaxy's disk and checks to see if any other particles' cross-sections 
($\pi r_{\rm{cld}}^2$) cross the vector. If there are none, then the gas 
particle is unshielded and will feel the acceleration $a_{\rm{ram}}$. In 
this case, an additional acceleration vector of magnitude $a_{\rm{ram}}$ 
is added to the particle's equation of motion, in the direction of the 
ICM wind.

For simplicity, we consider only the atomic gas content (HI) in our models, neglecting the presence of molecular gas (H$_2$), and Helium (He).
As H$_2$ clouds have a small cross-section, they are not expected to be efficiently ram pressure stripped (\citealp{Quilis2000})\footnote{Although see \cite{Vollmer2008} for possible stripped H$_2$.}. Thus in effect we have neglected an unstrippable mass component that may be $\sim$20$\%$ the mass of the HI and He combined (\citealp{Braine2001}). Including such a mass component could cause our model galaxies to respond as if they have gas fractions which are $\sim$10-15$\%$ lower than stated. However this effect may be entirely cancelled by our models not including a Helium component, which means we may actually underestimate the gas mass loss by as much as 40$\%$ (\citealp{Tielens2010}).

We explore models where the ram pressure wind encounters the TDG disk 
face-on. We do not expect our results are sensitive to this, as ram 
pressure is expected to affect disk galaxies in a similar manner for a 
wide range of wind inclination, except for near edge-on ram pressure 
stripping (\citealp{Vollmer2001}; \citealp{Marcolini2003}; 
\citealp{Roediger2006}).  We fix a constant velocity for the test galaxy 
and for the density of the ICM it is moving through. In the 
frame-of-reference of the test galaxy, it experiences an oncoming, 
uniform density, constant velocity wind. We refer to such tests as 
`wind-tunnel' tests. It should be noted that this situation is 
artificial for real galaxies, in clusters or groups, or in interacting 
galaxies, whose orbits subject them to both changing wind velocities and 
densities of hot gas. However, as we will later show, some aspects of 
the TDGs' response to ram pressure is very complex, even in these 
idealised wind-tunnel tests. We will consider time-varying ram pressures 
in a later study.

We choose an arbitrary fixed ICM density of $\rho_{\rm{ICM}}=10^{-4}$ Hydrogen atoms cm$^{-3}$. For a Virgo-like ICM distributed in a Beta-model (like 
cluster model `C1' of \citealp{Roediger2007}), our choice corresponds to 
densities found in the outskirts of the Virgo cluster (R$\sim$1000~kpc). 
This choice of density is also a reasonable approximation for the hot 
gas halo of the Milky Way (\citealp{Weiner1996}; 
\citealp{Stanimirovic2002}; \citealp{Putman2003b}; 
\citealp{Bregman2007}; \citealp{Lehner2011}; \citealp{Gupta2012}), or 
loose groups of galaxies (\citealp{Mushotzky2004}).

As we have fixed the ICM density, we vary the strength of ram pressure 
between wind tunnel tests, by choosing the wind speed $v_{\rm{wind}}$. 
We subject our standard galaxy model to an ICM wind of constant velocity 
$v_{\rm{wind}}$ with a face-on inclination; $v_{\rm{wind}}$ is varied 
from 100$-$800 km s$^{-1}$. This covers typical wind-velocities of dwarf 
galaxies orbiting in the Milky Way (100$-$200~km~s$^{-1}$), galaxies in 
loose groups (200$-$500~km~s$^{-1}$; \citealp{Zabludoff1998}), and the 
velocities of cluster galaxies ($\sim$800~km~s$^{-1}$).

As ram pressure is proportional to the intracluster medium density times 
the galaxy velocity squared, the ram pressures we consider are also 
representative of ram pressure in denser environments (such as cluster 
cores), or lower density environments (such as the outer halo of the 
Milky Way), if we rescale the wind speeds quoted in this paper by a 
constant factor. For example, for cluster model `C1' of 
\cite{Roediger2007} at a radius of $r\sim$250~kpc (near the cluster 
centre), the hot gas density is a factor $\sim$10 times higher than our 
choice of $\rho_{\rm{ICM}}$. However, equivalent ram pressures as those 
in this study, will occur for galaxies moving through such a medium if 
they have velocities $\sqrt{10}\approx$3 times {\it{lower}} than our 
stated velocities. Similarly, the hot gas density in the outer hot gas 
halo ($r$=250~kpc) of the Milky Way is a factor $\sim$10 times lower 
than our choice of $\rho_{\rm{ICM}}$ (\citealp{Blitz2000}; 
\citealp{Sembach2003}), but equivalent ram pressures will occur for 
galaxies with velocities approximately 3 times {\it{higher}}. For a 
complete list of the wind-tunnel tests applied to each of the models, 
see Table~\ref{windtest}.

Each model galaxy is subjected to the ICM wind for 2.5~Gyr. This 
duration is chosen as it allows sufficient time for gas that has been 
unbound to be accelerated away from the stellar disk. The duration is 
also physically motivated. \cite{Trentham2002} give the crossing time of 
the Virgo cluster as one tenth of a Hubble time. Therefore 2.5~Gyr 
represents a rough timescale for which a Virgo cluster galaxy might 
experience ram pressure over 1-2 orbits in the cluster.

\begin{figure*} \centering \epsfysize=12.5cm \epsffile{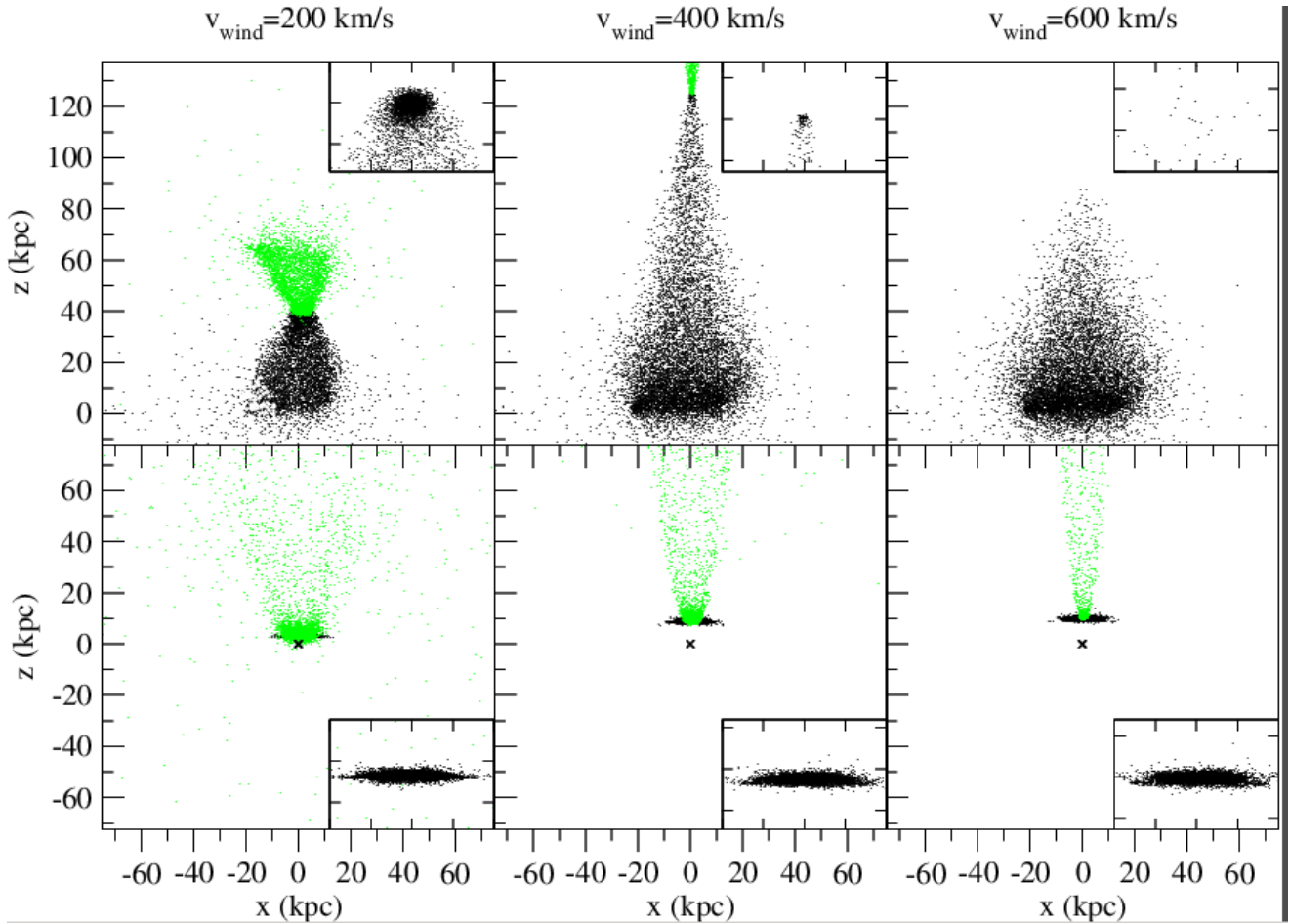} 
\caption{The impact of ram pressure stripping on the stars (black 
points) and gas (green points) of a model dwarf galaxy (M8gf70R2) in a 
150~kpc sided box. In the lower panels the model has a dark matter halo 
(like a typical dIrr). In the upper panels there is no dark matter halo 
(like a TDG). Ram pressure wind speeds of $v_{\rm{wind}}$=200, 400, and 
600 km~s$^{-1}$ (left, middle, and right columns) are presented. A 
`close-up' image of the stars only is shown in a corner sub panel 
spanning 20.0 by 13.3~kpc. Without a dark matter halo (upper panels), 
the effects of ram pressure are dramatic. Ram pressure stripping can 
cause significant stellar losses. The galaxy's stellar disk is also 
accelerated by ram pressure drag acting on the remaining gas disk, 
causing stars to lie in a stream pointing away from the gas stream. If 
the gas is rapidly stripped altogether (e.g., 
$v_{\rm{wind}}$=600~km~s$^{-1}$), there is little dragging, but the TDG 
is completely unbound. With a dark matter halo (lower panel), the dwarf 
galaxy is much more stable to ram pressure stripping. There are no 
stellar losses. Ram pressure results in much weaker acceleration of the 
stellar disk. The stellar disk merely thickens by a factor $\sim$2 in 
response to the loss of the gas mass.}
\label{yzplots}
\end{figure*}

We emphasise that this model should be regarded as a toy model of ram 
pressure. Recent ram pressure stripping models discussed in the 
literature (see for example \citealp{Roediger2007}) have advanced 
significantly beyond the simple ram pressure recipe of 
\cite{Vollmer2001}. Increasingly high resolution along the ISM-ICM 
boundary has allowed quantification of additional stripping mechanisms 
such as Rayleigh-Taylor (RT) instabilities and Kelvin-Helmholtz (KH) 
stripping. The ICM gas is normally composed of a live gas component that 
can form a shock front when a galaxy reaches supersonic velocities. 
Typically these studies concentrate on the highest possible resolution 
of the gas component of their disk galaxies, using analytical static 
potentials to treat the gravitational influence of the dark matter halo 
and stellar disk.

Our toy model does not include a live ICM gas component, so physically 
does not include the effects of RT instabilities or KH stripping. 
However, the toy model presented is fast, enabling wider parameter 
searches to be conducted and allowing us to include a live stellar 
component in our galaxy models. This is crucial to the results of this 
study. Furthermore, our main results arise due to the loss of the gas 
mass. If similar masses of gas are removed -- independent of the ISM 
treatment -- we expect similar effects as seen here. Therefore, we don't 
expect our key results to change substantially if we repeated our tests 
with a significantly more sophisticated treatment of the ISM. 
Furthermore, as was demonstrated in \cite{Smith2012}, despite its 
simplicity, the toy model can reasonably reproduce the evolution of the 
HI disk truncation radius in a similar manner as seen in much more 
complex ram pressure simulations.

\begin{figure*}
\centering \epsfysize=8.5cm \epsffile{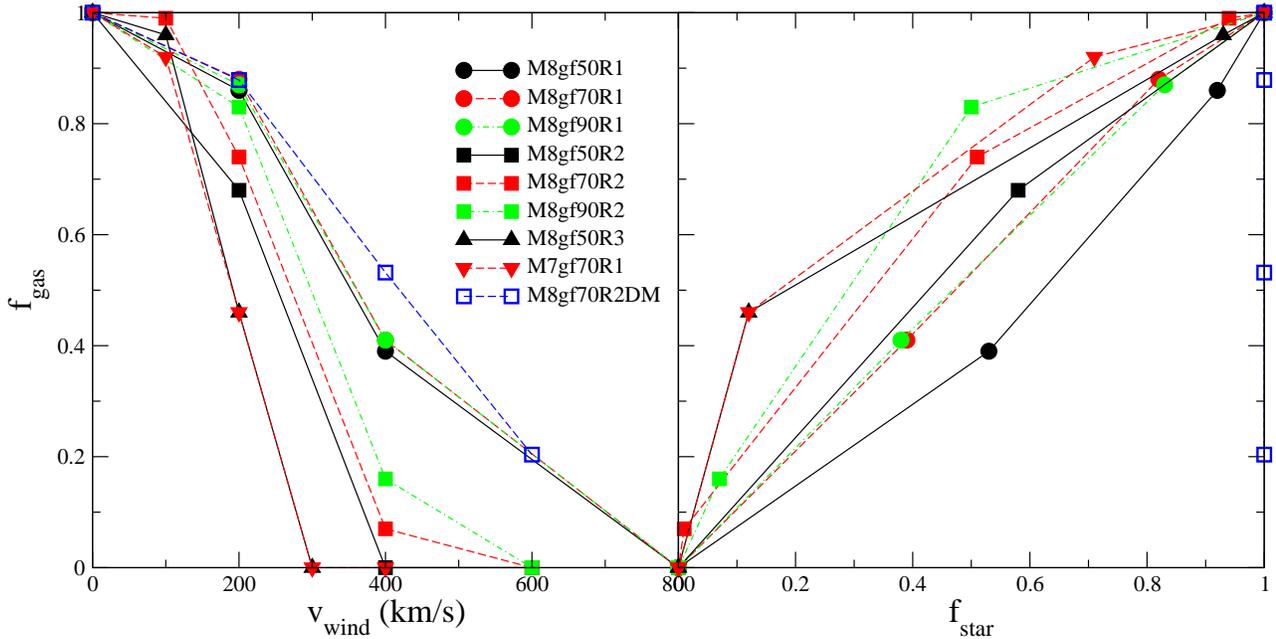}
\caption{(left) Fraction of gas remaining bound $f_{\rm{gas}}$ after wind 
tunnel tests with different wind speeds $v_{\rm{wind}}$. Symbol and line 
style indicates model galaxy. Models with lower central surface density 
(i.e large effective radius or low total mass) are stripped for lower wind 
speeds. (right) Fraction of gas $f_{\rm{gas}}$ that remains bound compared 
to the fraction of stars that remain bound $f_{\rm{star}}$. In general, 
the amount of gas that is stripped is roughly equal to the amount of stars 
that become unbound for our model TDGs. As the stars and gas are 
distributed in the same manner, this reflects the fact that the stars are 
unbound at the same radii where the gas is stripped. TDG Model M8gf70R2 (red squares) is the same as dIrr Model M8gf70R2DM (blue empty squares) except the latter has a dark matter halo. No stars are lost in the model with a dark matter halo. Also gas losses are reduced in the model with a halo, in part from the increased gravitational restoring force of the galaxy.}
\label{baryonlosses}
\end{figure*}

\section{Results}
\subsection{Effects of ram pressure enhanced in absence of dark matter halo}

TDGs may be very sensitive to ram pressure, as they are not expected to 
have a protective dark matter halo. To test this, we compare two dwarf 
galaxy models - one with a dark matter halo (we shall refer to this as the 
dIrr model), and an identical model but without a dark matter halo 
(referred to as the TDG model). The disks of both models have the 
properties of Model M8gf70R2, which are typical of TDGs.

We test the response of both models to three wind tunnel tests with wind 
speed $v_{\rm{wind}}$=200, 400, and 600~km~s$^{-1}$. After 2.5~Gyr of the 
wind tunnel test, we show the distributions of star (black points) and gas 
(green points) particles in Fig. \ref{yzplots}. Disks are viewed edge-on 
in a 150~kpc box, although a close-up view of the stars alone is provided 
in the accompanying 20$\times$13~kpc inset panels. The ram pressure wind 
flows upwards in each panel(in the positive z-direction).

We first consider the effects on the TDG model (upper row). Stripped gas 
is carried away in the direction of the ram pressure wind. For ram 
pressure winds with $v_{\rm{wind}}$=200 or 400~km~s$^{-1}$, the gas is not 
completely removed. Therefore, a truncated gas disk continues to provide a 
cross-section to the ram pressure wind, and so feels a drag force. The gas 
disk is accelerated by this force in the z-direction. As it is 
gravitationally bound to the stellar disk, the stellar disk also feels the 
drag force, and is similarly accelerated. However, this dragging mechanism 
can only operate efficiently on the stellar disk while gas remains in the 
disk. So, in the $v_{\rm{wind}}$=600~km~s$^{-1}$ case, which quickly loses 
all its gas, there is little dragging.

As gas is stripped, the loss of the gas mass causes stars to be unbound. 
In fact, where the gas is stripped, reflects where the stars are lost. For 
example, if only the outer disk gas is stripped (e.g., 
$v_{\rm{wind}}$=200~km~s$^{-1}$), only the outer disk stars are unbound, 
truncating the stellar disk in a similar way to that of the gas disk. 
However, if all the gas is stripped (e.g. 
$v_{\rm{wind}}$=600~km~s$^{-1}$), then all the stars are unbound -- 
{\it{the TDG is completely destroyed.}}

The final location of the stripped stars is dependent upon whether models 
are ram pressure stripped of all their gas, or only partially stripped. 
Those models that are only partially stripped (e.g., $v_{\rm{wind}}$=200 
and 400~km~s$^{-1}$), suffer simultaneous unbinding of their stars and 
acceleration of their disk in the direction of flow of the ram pressure 
wind. The net result of star losses and disk acceleration is that unbound 
stars tend to lie in streams. Furthermore the stellar streams point in 
opposite directions to the gas streams in our wind tunnel tests. This is 
somewhat at odds with the behaviour of tidally stripped streams of stars 
which are more closely aligned with the gas streams \citep{Connors2006}. 
However, the extreme linear shape of the gas and star streams we see is 
almost certainly an artifact of our use of wind tunnel tests, where no 
external potential affects their dynamics -- unlike in real TDGs.

However, models that lose all their gas very quickly (e.g., 
$v_{\rm{wind}}$=600~km~s$^{-1}$) suffer little dragging, and so unbound 
stars expand away from the TDGs centre in a roughly isotropic manner, 
creating an expanding envelope.

Now we compare the effects of the wind tunnel tests on the TDG model to 
the dIrr model that has a dark matter halo (in the lower panels of Fig. 
\ref{yzplots}). The dIrr model also suffers ram pressure drag, however the 
resulting acceleration is much less significant. We indicate the initial 
position of the disk with a cross symbol. The gas disk continues to feel a 
drag force, like in the TDG model, but now the drag force must be shared 
between the the stellar disk {\it{and}} the dark matter surrounding the 
disk. As the additional mass of the dark matter must be towed, the 
resulting acceleration is much weaker.

There is no unbinding of stars in the dIrr model, and hence no stellar 
streams. The loss of the gas mass causes a much less dramatic effect on 
the stellar dynamics. This is because the dark matter potential plays a 
strong role in influencing the stellar dynamics in the dIrr model. The 
response of the stars is limited to a thickening of the stellar disk by a 
factor $\sim$2, due to the presence of the dark matter. We note that a 
thorough investigation of the effects of ram pressure stripping on dark 
matter dominated dwarfs can be found in \cite{Smith2012}.

In summary, due to an absence of dark matter in our model TDG, the effects 
of ram pressure are very dramatic. Although ram pressure only directly 
impacts upon the gas component of the TDG, the loss of the gas can have a 
very substantial effect on its stellar disk. Stars are unbound, the disk 
is accelerated, and the entire TDG is destroyed when the gas is fully 
stripped.

\subsection{Tidal Dwarf galaxy disruption through ram pressure}

We have so far only considered a single TDG model -- Model M8gf70R2. In 
the following section we better quantify the loss of gas and stars for 
TDGs with a variety of properties, each experiencing different strengths 
of ram pressure. A full list of the model TDGs we consider, and the speed 
of the wind in each wind tunnel test, is given in Table~\ref{windtest}.

We record the final bound fractions of stars $f_{\rm{star}}$, and gas 
$f_{\rm{gas}}$, for each model, after each wind tunnel test at t=2.5~Gyr. To measure if a star or gas particle is bound to the TDG, we use the `snowballing' method described in \cite{Smith2013}. The results 
are shown in Fig. \ref{baryonlosses}. The left panel shows the dependency 
of $f_{\rm{gas}}$ on the ram pressure wind speed, for each model TDG (see 
the key). As expected, stronger wind speeds causes more gas to be 
stripped. Those models with lower central surface densities are most 
sensitive to gas stripping by ram pressure. These include models with 
larger effective radii (e.g M8gf50R3), or lower mass (e.g. M7gf70R1), 
which lose about half their gas at $v_{\rm{wind}}$=200~km~s$^{-1}$, and 
are entirely stripped of their gas at $v_{\rm{wind}}$=300~km~s$^{-1}$. Ram 
pressure winds with $v_{\rm{wind}}$$\sim$200~km~s$^{-1}$ are expected for 
some dwarf galaxies bound to the MW, while winds on the order of 
$v_{\rm{wind}}\ge$300~km~s$^{-1}$ are expected in group environments. 
Higher surface density models, such as those with an effective radius of 
1~kpc (M8gf50R1-M8gf90R1), lose roughly half their gas at 
$v_{\rm{wind}}$=400~km~s$^{-1}$ (galaxy group velocities), and are not 
completely stripped until $v_{\rm{wind}}$=800~km~s$^{-1}$ (galaxy cluster 
velocities).

\begin{figure}
\centering \epsfysize=6.5cm \epsffile{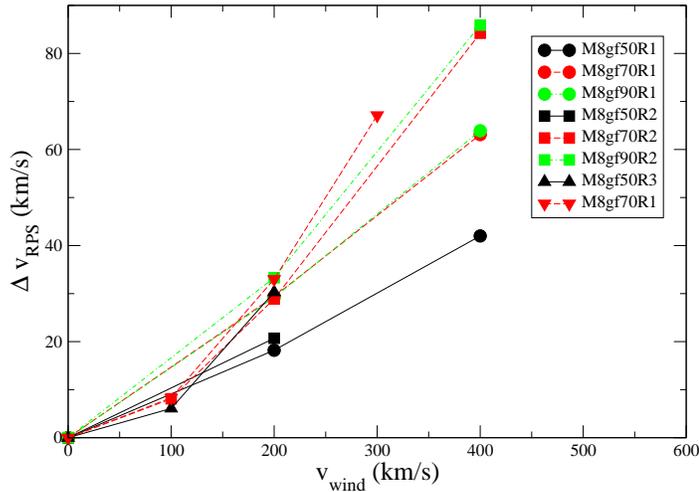}
\caption{The gas disk experiences a drag force which is shared with the 
stellar disk through their gravitational interaction. We measure the 
change in velocity of the bound stellar disk of the model as a result of 
ram pressure drag $\Delta V_{\rm{RPS}}$ and plot it versus the wind speed 
$v_{\rm{wind}}$ in the wind tunnel test. Changes in velocity by 
20-80~km~s$^{-1}$ are typical. Most lines end abruptly at higher ram 
pressure winds when models are entirely stripped of their gas and their 
stellar disk is destroyed, so as the change in velocity of the bound 
stellar disk can no longer be measured.}
\label{deltavrps}
\end{figure}

The right panel of Fig.~2 allows us to compare $f_{\rm{gas}}$ to 
$f_{\rm{star}}$ for all simulations in our parameter study. In general, we 
see that almost all models suffer roughly equal fractions of stars and gas 
being stripped. For example, if a model has half its gas stripped, then 
roughly half its stars are unbound in the process. This corroborates the 
point made earlier that stars are unbound where the gas is lost, as the 
gas is initially distributed in a disk with the same scalelength as the 
stars. Therefore, if the gas was more extended than the stars, then we 
might expect preferential loss of gas over stars \citep{Connors2006}.

In the previous section, we compared the difference in the appearance of Model M8gf70R2 after undergoing ram pressure with, and without, a dark matter halo surrounding the disk (see Fig. \ref{yzplots}). Now we include the model with a halo in Fig. \ref{baryonlosses} so we can investigate how baryonic mass loss depends on the presence of a halo. The dark matter dominated model is referred to as Model M8gf70R2DM (open blue squares), whereas the dark matter free TDG model is Model M8gf70R2 (filled red squares). The model with a dark matter halo suffers no stellar losses at all, even when a large fraction of the gas is stripped. We also see that less gas is stripped in the model with a dark matter halo. This occurs partly due to the enhanced gravitational restoring force from the dark matter halo (\citealp{Abadi1999}). Also, when there is no dark matter halo surrounding the disk, the stars near the truncation radius maybe highly perturbed by the gas removal. This may also play a role in enhancing gas losses in TDGs.

\subsection{Stellar disk acceleration by ram pressure drag, and enhanced 
stellar losses}

We now investigate the degree of acceleration that the stellar disk of a 
TDG undergoing ram pressure might experience. Therefore, we record the 
total change in velocity that occurs to the stellar disk as a result of 
ram pressure $\Delta V_{\rm{RPS}}$, after 2.5~Gyr of each wind tunnel 
test. We measure $\Delta V_{\rm{RPS}}$ for all the tests in our parameter 
study, to try to understand any dependency on TDG properties.

The results are shown in Fig. \ref{deltavrps} for each model TDG (see 
symbols and line styles in the accompanying key). Most models show a total 
change in velocity due to ram pressure drag of 20-80~km~s$^{-1}$. We 
measure the velocity of the bound stars, therefore most models have lines 
that abruptly end at an upper $v_{\rm{wind}}$ when their gas is stripped, 
and the stars become unbound. This complicates the clear detection of a 
simple relationship between amount of acceleration and TDG properties, as 
it is convolved with a dependency on survival to ram pressure. As a 
result, we do not see a trivial dependency on model properties. However, we 
emphasise that a change in velocity of $\sim$80~km~s$^{-1}$ may be 
significant for altering orbital properties of TDGs, in particular where 
the velocity change is a significant fraction of the galaxy's orbital 
velocity. As we will now demonstrate, the change in velocity of the TDG 
disk also has consequences for the number of stars that are unbound.

To demonstrate that stars are lost due to acceleration of the TDG, we 
consider a low gas fraction model (30$\%$ gas, 70$\%$ stars): M8gf30R2. We 
submit this model to four ram pressure wind tunnel tests for 2.5~Gyr 
($v_{\rm{wind}}$=200, 300, 400 and 600~km~s$^{-1}$) and measure the final 
bound fraction of stars $f_{\rm{star}}$, and gas $f_{\rm{gas}}$. We also 
test the resulting $f_{\rm{star}}$, allowing for instantaneous removal of 
the gas. This `instant gas mass loss' case allows us to see the effects of 
the removal of the gas mass without the additional effects, such as ram 
pressure dragging, that we see in the wind-tunnel test.

\begin{figure}
\centering \epsfysize=6.5cm \epsffile{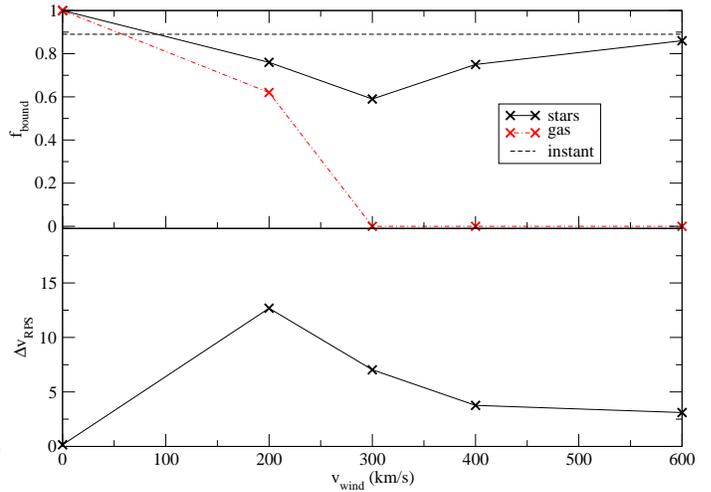}
\caption{For model M8gf30R2; (upper) fraction of stars $f_{\rm{star}}$ 
(black solid line), and fraction of gas $f_{\rm{gas}}$ (red dash-dot curve) 
as a function of wind speed in each wind tunnel test. The dashed 
horizontal line indicates $f_{\rm{star}}$ after instantaneous removal of 
the gas. (lower panel) Change in velocity of the stellar disk of the model 
as a result of ram pressure drag $\Delta V_{\rm{RPS}}$. Stars are unbound 
due to the loss of the gas potential. However, acceleration of the stellar 
disk by ram pressure drag, further enhances stellar losses. For example 
(see upper panel) ram pressure stripped models experiences greater stellar 
losses than occur when the gas is instantaneously removed. Also, if the 
gas is totally stripped (see lower panel), models that have experienced 
the greatest acceleration due to ram pressure suffer larger stellar 
losses.}
\label{draglosses}
\end{figure}

This latter case is shown in Fig. \ref{draglosses}. The upper panel shows 
the final $f_{\rm{star}}$ (black line) and $f_{\rm{gas}}$ (red line) of 
the TDG model in the wind tunnel tests with wind speed $v_{\rm{wind}}$ 
(shown on the x-axis). The dashed horizontal line indicates the final 
$f_{\rm{gas}}$ of the model after instantaneous gas loss. In all the wind 
tunnel tests, the model loses {\it{more}} stars than in the instantaneous 
gas loss case.

The latter is somewhat surprising, as the instant gas loss case sets an 
upper limit on the number of stars that are unbound by the loss of the gas 
potential alone. Instant gas loss gives no time for the stars to respond 
to the changing potential, whereas the loss of the gas is not so quick in 
our wind tunnel tests, so the stars have more time to respond to the 
changing potential. In fact, even more surprising is the wind tunnel test 
with $v_{\rm{wind}}$=200~km~s$^{-1}$ -- {\it{the model loses less than 
half its gas, and still more stars are unbound than when all the gas is 
lost instantaneously.}}

\begin{figure*}
\centering \epsfysize=5.2cm \epsffile{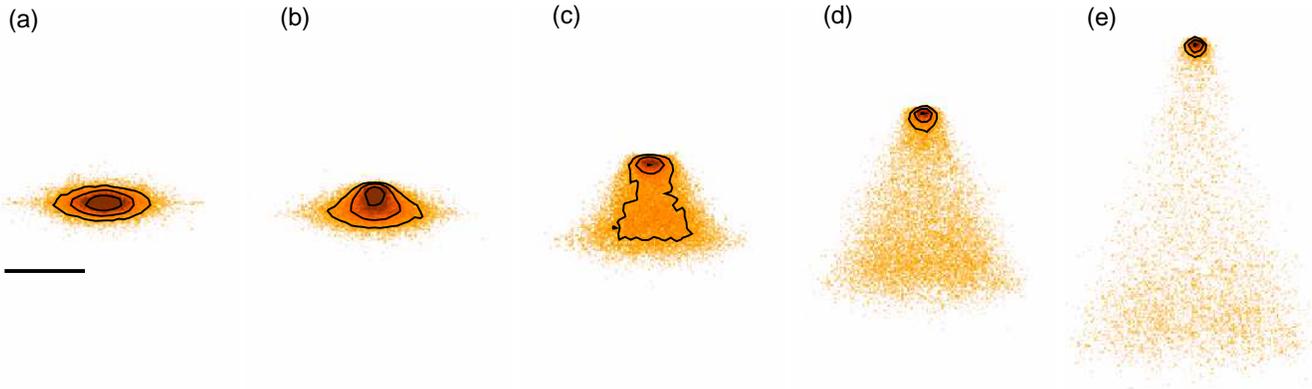}
\caption{Snapshots of surface brightness time-evolution of Model M8gf50R3 
at time = (a) 0.000~Gyr, (b) 0.625~Gyr, (c) 1.250~Gyr, (d) 1.875~Gyr, (e) 
2.500~Gyr. Length of solid bold line on left of figure indicates scale of 
10~kpc. Intensity of colour indicates surface brightness, with contours 
shown at 29, 30, 31 mag$_{\rm{V}}$~arcsec$^{-2}$. For simplicity we assume 
a stellar mass-to-light ratio of 1 in all surface brightness calculations 
(although this may be very incorrect for a young stellar population).}
\label{streamSB}
\end{figure*}

These results highlight the presence of an additional mechanism that 
causes stars to be unbound, and in greater quantities than can occur from 
the loss of the gas potential alone. We find that the additional mechanism 
is due to acceleration of the TDG disk by ram pressure drag. This can be 
understood if we consider the accelerating frame of reference of a TDG 
undergoing ram pressure drag. Stars are liberated from the disk by the 
loss of gas mass. If the TDG were not accelerating, some stars would 
remain marginally bound and eventually infall back towards the disk. 
However, due to the acceleration of the TDG since the stars were emitted, 
these stars are left behind. This is apparent when we plot the change in 
velocity of the stellar disk as a result of ram pressure $\Delta 
V_{\rm{RPS}}$ in the lower panel of Fig. \ref{draglosses}. When the gas is 
fully stripped ($v_{\rm{wind}}\ge$200~km~s$^{-1}$), models that accelerate 
the most (i.e with the highest values of $\Delta V_{\rm{RPS}}$) lose more 
stars. We note that as $v_{\rm{wind}}$ increases from 300~km~s$^{-1}$ to 600~km~s$^{-1}$, the fraction of bound stars approaches the instantaneous gas expulsion value. This is to be expected as an increasing $v_{\rm{wind}}$ leads to an increasingly rapid total removal of the gas.

We have demonstrated the effects of disk acceleration on $f_{\rm{star}}$ 
with a (somewhat unrealistic) low gas fraction model, simply because 
instantaneous gas loss results in the destruction of all the other models, 
therefore hiding the effects of the disk acceleration. However, loss of 
stars due to gas mass loss, combined with acceleration due to ram pressure 
drag, undoubtedly occurs in all our ram pressure stripped models.

\subsection{Formation of stellar streams by ram pressure}

Stellar losses, combined with acceleration by ram pressure drag, causes 
stars to lie in a low surface brightness stream behind the accelerated 
TDG. In Fig. \ref{streamSB} we show the stream produced by Model M8gf50R3. 
Each image is a snapshot at 0.625~Gyr intervals, ranging from $t$=0~Gyr to 
$t$=2.5~Gyr. The bold line indicates a 10~kpc scale. The intensity is 
proportional to the logarithm of the mass surface density. We include a 
surface brightness contour at 29, 30, 31 mag$_{\rm{V}}$~arcsec$^{-2}$, 
assuming a V-band stellar mass-to-light ratio ($M/L_{\rm{V}}$) of 1. We 
note that this choice of stellar mass-to-light ratio may be incorrect, as 
unbound stars may consist of a complex mix of young and old stellar 
populations. However, as we do not include star formation in our models, 
we cannot provide detailed predictions of the stellar populations in the 
streams, and for simplicity choose a single fixed value of 
unity.\footnote{If we assume a younger, brighter population with 
$M/L_{\rm{V}}$=0.25, the value associated with each surface brightness 
contour decreases by one magnitude.}

By $t$=1.25~Gyr, an $\sim$15~kpc stream has been drawn out. By 2.5~Gyr the 
stream is $\sim$40~kpc long. With increasing length, the stream decreases 
in surface brightness. What is apparent is that the long streams are of 
very low surface brightness. Even the relatively bright stream at 
$t$=1.25~Gyr (panel c), is only seen at $>$31 mag$_{\rm{V}}$~arcsec$^{-2}$ 
(assuming $M/L_{\rm{V}}$=1). Such low surface brightness streams may only be detected surrounding Local Group galaxies, and even then only from star counts. For this stream to be visible at the level of 
29~mag$_{\rm{V}}$~arcsec$^{-2}$, the stellar population must be $\sim$6 
times brighter than we have assumed. However the situation at 
$t$=0.625~Gyr (panel b) may be more easily detected, when the core of the 
galaxy is several kiloparsecs off-centre within a surrounding low surface 
brightness envelope.

\subsection{Effects on mass Sersic profiles}

\begin{figure*}
\centering \epsfysize=22.0cm \epsffile{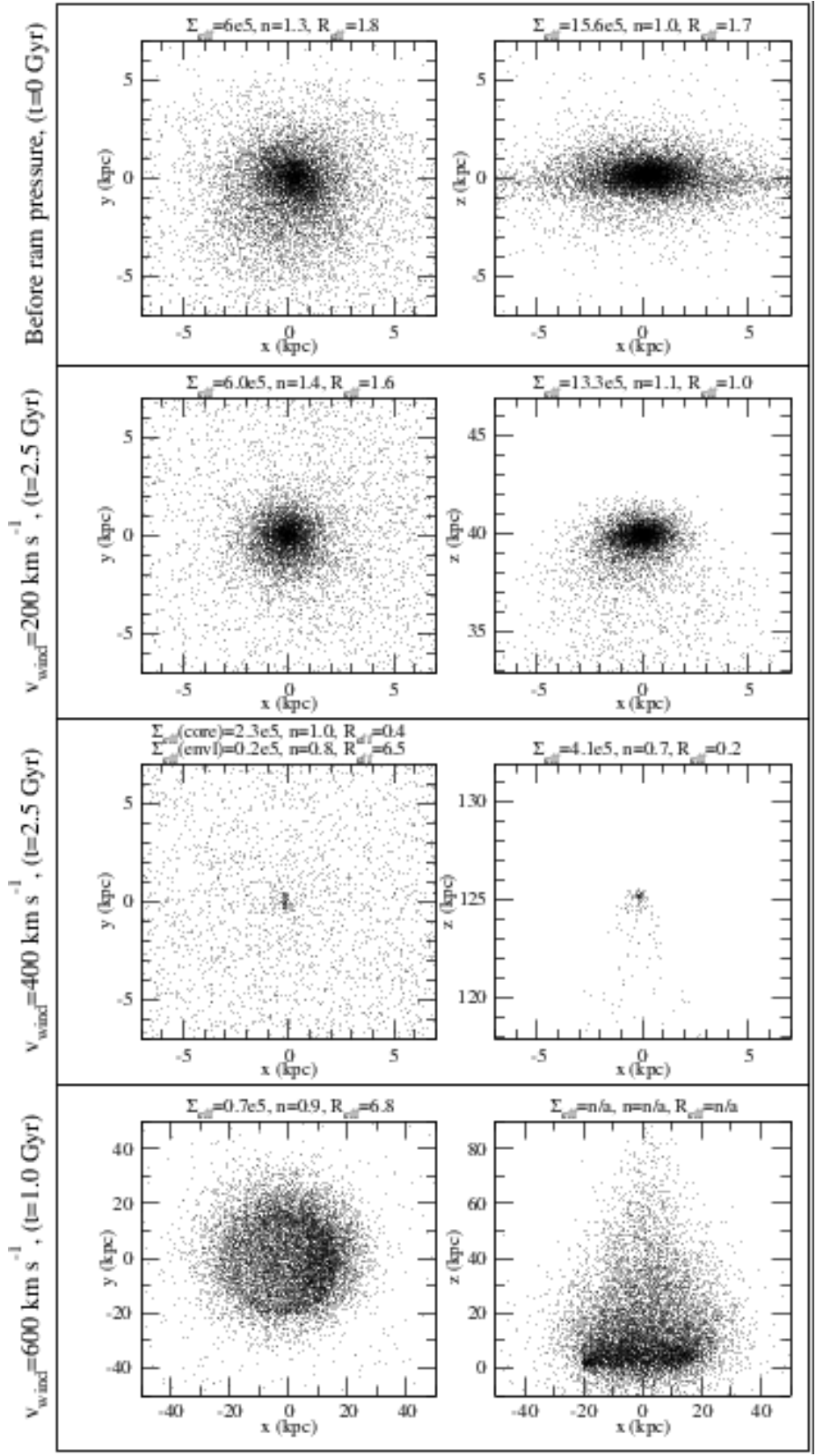}
\caption{Face-on (left column) and edge-on (right column) snapshots of the 
star particle distribution of Model M8gf70R2 following wind tunnel tests 
with wind speed $v_{\rm{wind}}$=200, 400, 600~km~s$^{-1}$ (second, third, 
and fourth row respectively). Above each panel, the parameters of the 
Sersic fit to the mass distribution are shown; surface density within the 
effective radius ($\Sigma_{\rm{eff}}$), the Sersic index (n), and the 
effective radius (R$_{\rm{eff}}$). In general, the Sersic index remains 
near n$\sim$1. However R$_{\rm{eff}}$ can be reduced by preferential loss 
of outer disk stars in TDGs that are not completely stripped of their gas. 
Or if the gas is rapidly stripped, the expanding unbound stars can have 
very large R$_{\rm{eff}}$, but coupled with low $\Sigma_{\rm{eff}}$.}
\label{sersicprofiles}
\end{figure*}

We now examine the effects of stellar losses on the stellar 
surface density profiles. We restrict ourselves to studying the evolution 
of the {\it{mass}} surface density, and not the luminosity. In Fig. 
\ref{sersicprofiles}, we show the stellar particle distribution of Model 
M8gf70R2 before ram pressure (first row), and after undergoing three wind 
tunnel tests: $v_{\rm{wind}}$=200, 400, 600~km~s$^{-1}$ (second, third, 
and fourth row respectively). The left and right column shows the TDG 
viewed face-on and edge-on respectively.

We produce fits images of the stellar mass distribution using the {\sc{IRAF}} {\it{rtextimage}} task. Our fits images cover a 20 by 20~kpc region for the pre-ram pressure model, and $v_{\rm{wind}}$=200 and 400~km~s$^{-1}$ models. We use a 40 by 40~kpc region fits image for the $v_{\rm{wind}}$=600~km~s$^{-1}$ model. We measure the surface density profile of the stars using the {\sc{IRAF}} {\it{ellipse}} task, without restricting the radial range to be fitted. The resulting mass surface density profile of the stars is then fitted over the complete radial range using a generalised Sersic profile (\citealp{Caon1993}):

\begin{equation}
\Sigma(R)=\Sigma_{\rm{eff}} \exp \left(-b_{\rm{n}} \left[ \left(\frac{R}{R_{\rm{eff}}}\right)^{1/n}-1 \right] \right)
\end{equation}
\noindent
where $b_{\rm{n}}=1.9992n-0.3271$, $n$ is the Sersic index, 
$R_{\rm{eff}}$ is the effective radius, and $\Sigma_{\rm{eff}}$ is the 
surface density at $R_{\rm{eff}}$. The best-fit values are shown above 
each panel in Fig.~\ref{sersicprofiles}.

\begin{figure}
\centering \epsfysize=11.0cm \epsffile{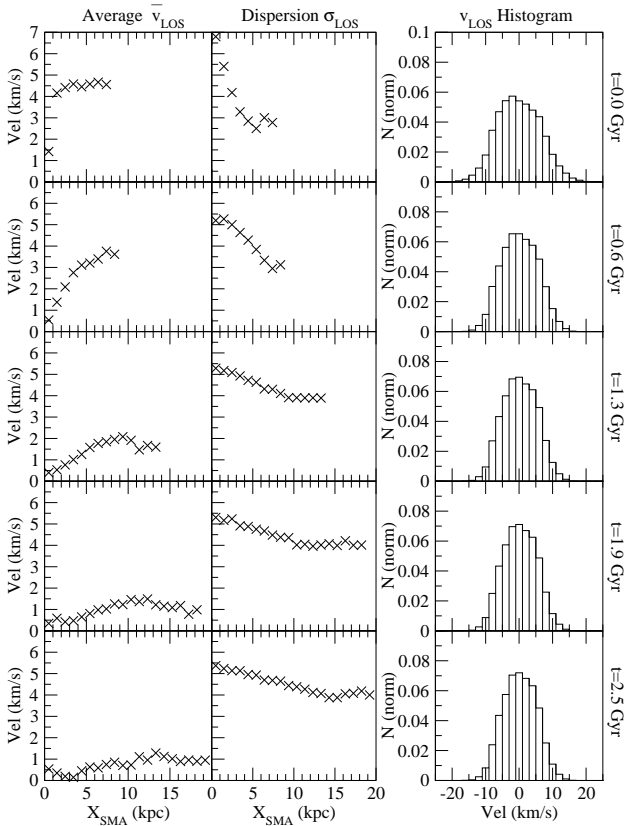}
\caption{Plots of the line-of-sight velocities of Model M8gf70R2, viewed along the y-axis so as the disk is seen edge-on. In column 1 and 2, velocities are binned along the semi-major axis of M8gf70R2. Column 1 is the average 
velocity in each bin (so indicates rotation). Column 2 is the velocity dispersion in each bin. Column 3 is a histogram of the line-of-sight velocities of all the star particles. Each row is from a specific time (snapshot time is shown along the right-hand edge of Column 3). The model undergoes a wind 
tunnel test with $v_{\rm{wind}}$=600~km~s$^{-1}$. All the gas is stripped, 
causing the stars to be totally unbound, by t=0.6~Gyr. The peak of the 
rotation curve decreases, and the velocity dispersion profile flattens as 
the unbound stars expand. However, if viewed at only one instant, there is no clear indication that the stars are actually unbound and expanding, in either the radial plots or the velocity histograms.}
\label{dynamics}
\end{figure}

Prior to ram pressure (upper row), the stellar distribution is close to 
exponential (n$\sim$1), and the effective radius is approximately 2~kpc. 
However, after the 200~km~s$^{-1}$ wind tunnel test (second row), the 
outer disk gas has been stripped, resulting in unbinding of the outer disk 
stars. This truncates the stellar disk, in the same way as the gas disk 
has been truncated, at a radius $R$$\sim$3~kpc. The stellar disk inside the 
truncation radius remains roughly exponential. However, the loss of the 
outer disk stars causes the effective radius to be reduced.

A similar but more extreme example is seen in the 400~km~s$^{-1}$ wind 
tunnel test (third row). Now the gas and stellar disks are heavily 
truncated at $R<$0.5~kpc. Once again, the remaining stellar disk is 
roughly exponential, but now the effective radius has been significantly 
reduced. Viewed face-on (left panel), the presence of large numbers of 
stripped stars down our line-of-sight causes a low surface density 
envelope to appear around the heavily truncated stellar disk. There is no attempt to make a cut in the z-direction, therefore the line-of-sight contains all the star particles possible. We fit the surface profile of the truncated stellar disk (or `core') separately from the surrounding envelope. This is accomplished by only fitting data points at $R>$1.0~kpc for the envelope, and points at $R<$0.5~kpc for the core. The 
envelope also has a roughly exponential profile ($n$$\sim$1), although the 
effective radius is very large ($\sim$6~kpc).

A rather different surface density profile occurs if stripping of all the 
gas occurs rapidly, such that there is little dragging. The loss of all 
the gas completely unbinds the stellar disk, and the lack of dragging 
means unbound stars do not form a stream. Instead, they expand outwards 
from the centre of the TDG in an envelope. This is seen in the 
600~km~s$^{-1}$ wind tunnel test (bottom row). As all the stars are 
unbound, we find the mass profile evolves rapidly with time. It becomes 
difficult to measure the parameters of the generalised Sersic profile 
beyond 1~Gyr. However, over $\sim$1~Gyr of evolution, the profile remains 
roughly exponential ($n\sim1$), becomes very low surface brightness, and 
increases its effective radius from $\sim$2~kpc to $\sim$6~kpc.

In summary, the surface density profiles of our ram pressure stripped 
models are all roughly exponential. However, if the TDG disk is not 
completely stripped of gas, stars in the outer disk are unbound where the 
gas has been lost. This causes the remaining stellar disk to be truncated, 
resulting in a reduced effective radius. On the contrary, if the gas is 
completely stripped from the TDG model, the stars are completely unbound, 
causing the effective radius to grow with time, while decreasing the 
surface brightness.

\subsection{Effects on stellar dynamics and implications for dynamical 
mass measurements}

We now examine the effects of ram pressure stripping on the dynamics of 
stars in our models. As stars are unbound when the gas is stripped, the 
dynamics of the unbound stars down our line-of-sight may potentially 
affect measurements of a TDG's dynamical mass from its stellar dynamics if 
dynamical equilibrium is assumed.

First we examine the stellar dynamics of the Model M8gf70R2. We study the 
time evolution of the stellar dynamics as viewed down a line-of-sight. We 
choose a sightline that is edge-on to the disk, as this best enables us to 
measure signatures of rotation. We arbitarily choose the y-axis as our line-of-sight, although our results show negligible change if we had instead chosen the x-axis. We bin the line-of-sight velocities of all stars along 
the semi-major axis of the disk, and calculate the average and standard 
deviation of the stellar velocities within that bin. Any trace of rotation 
should be visible in the profile of the average velocity.

The results for the wind tunnel test with $v_{\rm{wind}}$=600~km~s$^{-1}$ 
are shown in Fig. \ref{dynamics}. Recall from the previous section that 
this model is quickly stripped of all its gas, causing the stellar disk to 
become totally unbound. The unbound stars form an expanding envelope. Column 1 shows the average, and the Column 2 shows the standard 
deviation, of the velocities in a bin, plotted against the distance along 
the semi-major axis $X_{\rm{SMA}}$. We include all star particles that fall in a bin of $X_{\rm{SMA}}$ in the average and standard deviation calculation, without considering their z-position. Although we find negligible change in our results if we include only stars with $\lvert z \rvert<$10~kpc from the TDG centre. However, we neglect data points with less that 
50 stars in a bin to avoid low-N noise. 

The bins of average velocity (Column 1) show that a signature of rotation remains within the expanding 
envelope, although the peak value decreases. This shows the decreasing 
speed of rotation as the disk expands, as required for angular momentum 
conservation. Likewise, the velocity dispersion profile (Column 2)
becomes steadily flattened as the envelope expands.

Close to the core of the model ($X_{\rm{SMA}}<2$~kpc), one is clearly 
dominated by dispersion even before ram pressure, and this remains the 
case afterwards. However, prior to ram pressure, in the outer disk 
($X_{\rm{SMA}}>5$~kpc), rotation dominates dispersion. After ram pressure 
stripping, the dispersion in the outer disk tends to increase, while 
rotation tends to decrease. The net effect is a decrease in 
rotation-to-dispersion ratios, especially in the outer disk.

Interestingly, if we consider only one snapsot in time, there is no clear signature in either the average or 
standard deviation of the binned velocities indicating that the model is 
unbound and expanding. To look for another signature of the expansion, we plot histograms of the line-of-sight velocities of all the stars in a snapshot. These are shown in Column 3 of Fig. \ref{dynamics}. If all the stars were expanding towards and away from us down our line-of-sight with a roughly equal expansion velocity, we would see a double peak in the velocity histogram. In fact the stars expand with a smooth spread in 
velocities, so as there is no double peak of velocities. {\it{Therefore a TDG model that is completely and rapidly 
stripped of its gas shows very little indication that it has been entirely 
unbound if viewed at one instant.}} Recalling from the previous section, such a TDG remains near exponential, but steadily grows in size, 
decreasing its surface brightness while increasing its effective radius, 
and yet shows little indication of the fact it is unbound in its 
line-of-sight dynamics. 

Therefore, in some circumstances, ram pressure disrupted TDGs might be confused with other types of dwarf galaxies. With time the effective radius of our disrupted TDG model M8gf70R2 grows very large, and this might naively be expected to differentiate it from other dwarfs. However our effective radii are measured in the idealised situation when there is no background light, whereas in observations of real dwarfs the low surface brightness outer edges of the surface brightness profile may be lost in the background. If not correctly accounted for, this could result in measurements of the effective radius that are smaller than we measure in the models. Furthermore, our models do not consider the surrounding tidal field, that can preferentially truncate and strip the stars with the largest radii, physically reducing the effective radius. Also, we see no obvious reason why TDGs that start off much smaller, and are then disrupted by ram pressure, might not expand to have properties that overlap with those of other dwarf galaxies. The lack of dark matter, as seen in the stellar dynamics of a TDG, might also be expected to differentiate a TDG from another dwarf galaxy. However, as we will now demonstrate, the effect of ram pressure on the stellar dynamics of a TDG can mimic the presence of substantial quantities of dark matter.

In order to measure dynamical masses of galaxies, using their observed 
stellar dynamics, the stars are typically assumed to be in dynamical 
equilibrium with the galaxy's potential. However, our models show that ram 
pressure stripped TDGs have many stars unbound. The presence of unbound 
stars down our line-of-sight would violate the assumption of dynamical 
equilibrium, and may potentially result in dynamical masses that appear 
greater than the true mass. We attempt to quantify this effect now.

Following a similar approach as described in \cite{Evans2003} (and applied 
in \citealp{Beasley2006}; \citealp{Beasley2009}), we separate the total 
dynamical mass of a galaxy model $M_{\rm{dyn}}$ into a dynamical mass 
supported by pressure $M_{\rm{press}}$, and a dynamical mass supported by 
rotation $M_{\rm{rot}}$ such that 
$M_{\rm{dyn}}=M_{\rm{press}}+M_{\rm{rot}}$.

We calculate $M_{\rm{dyn}}$ at the half-mass radius $r_{\rm{1/2}}$. For 
the pressure supported dynamical mass, we use $M_{\rm{press}} 
(r<r_{\rm{1/2}}= 3~G^{-1} {<\sigma^2_{\rm{LOS}}}>r_{\rm{1/2}}$ 
(\citealp{Wolf2010}) where ${<\sigma^2_{\rm{LOS}}>}$ is the square of the 
line-of-sight velocity dispersion measured within $r_{\rm{eff}}$. This 
approach minimises the potential influence of velocity anisotropy on 
$M_{\rm{press}}$. We calculate the rotation-supported dynamical mass at 
$r_{\rm{1/2}}$ using $M_{\rm{rot}}=r_{\rm{1/2}}~G^{-1}~v_{\rm{rot}}^2$ 
where $v_{\rm{rot}}$ is the value of the average velocity, down our 
line-of-sight, measured at $r_{\rm{1/2}}$. Having calculated the total 
dynamical mass, we then normalise it by the true mass $M_{\rm{real}}$ also 
measured within $r_{\rm{1/2}}$. Specifically $M_{\rm{real}}$ is the total mass, bound or unbound, found within a sphere of radius $r_{\rm{1/2}}$. The ratio $M_{\rm{dyn}}/M_{\rm{real}}$ is 
a measure of how well the dynamical mass agrees with the true mass (e.g., 
if $M_{\rm{dyn}}/M_{\rm{real}}$$\equiv$1, the dynamical mass is in perfect 
agreement with the true mass).

As in the previous section, we consider Model M8gf70R2, evolved in 
isolation, and undergoing the same three wind tunnel tests: 
$v_{\rm{wind}}$=200, 400, and 600~km~s$^{-1}$. We consider an edge-on view 
to the disk (see right column of Fig.~\ref{sersicprofiles} for the final 
distribution of stars in each wind tunnel test along this line-of-sight). 
We follow the time evolution of the ratio $M_{\rm{dyn}}/M_{\rm{real}}$.

We first consider the model evolved for 2.5~Gyr in isolation. In 
isolation, we measure $M_{\rm{dyn}}/M_{\rm{real}}$=1.08, 0.90, 0.97, 0.93, 
0.98, and 1.02, when measured at $t$ = 0, 0.5, 1.0, 1.5, 2.0 and 2.5~Gyr, 
respectively. Thus, our technique for measuring the dynamical mass appears 
reliable to within $\sim$10\% of the true mass.

Next, we consider the $v_{\rm{wind}}$=600~km~s$^{-1}$ wind tunnel test, 
where the model was quickly stripped of all its gas, resulting in an 
expanding envelope of unbound stars. We measure 
$M_{\rm{dyn}}/M_{\rm{real}}$=1.1, 2.0, 6.1, 7.3, 8.0, 10.7, and 12.7, when 
measured at $t$ = 0.00, 0.25, 0.50, 0.63, 0.75, 1.00, and 1.25~Gyr, 
respectively. As expected, by assuming dynamical equilibrium, our 
technique finds dynamical masses that are heavily in excess of the real 
mass. The excess increases as the envelope expands.

Finally, we consider the wind tunnel tests with $v_{\rm{wind}}$=200 and 
400~km~s$^{-1}$. Recalling from the previous section, the model stellar 
disk suffers a mild disk truncation in the outer disk for 
$v_{\rm{wind}}$=200~km~s$^{-1}$ (see right panel of second row in Fig. 
\ref{sersicprofiles}). The disk truncation is much more severe for 
$v_{\rm{wind}}$=400~km~s$^{-1}$, and only a small stellar disk remains 
(see right panel of third row in Fig. \ref{sersicprofiles}). However, in 
both these cases, where a bound stellar disk remains, disk dragging has 
caused the unbound stars to be drawn into a stellar stream. Therefore 
there are fewer unbound stars seen down our line-of-sight, and their 
effect on mass estimates is greatly reduced. For 
$v_{\rm{wind}}$=200~km~s$^{-1}$, we measure 
$M_{\rm{dyn}}/M_{\rm{real}}$=1.02, 1.09, 1.10, 1.13, 1.22, and 1.29 at 
t=0.0, 0.5, 1.0, 1.5, 2.0, and 2.5~Gyr. For the more heavily truncated 
stellar disk with $v_{\rm{wind}}$=400~km~s$^{-1}$, we measure 
$M_{\rm{dyn}}/M_{\rm{real}}$=1.03, 1.69, 1.40, 1.35, 1.49, and 2.20 at 
$t$=0.0, 0.5, 1.0, 1.5, 2.0, and 2.5~Gyr.

In summary, the effect of ram pressure stripping on TDG stellar dynamics 
causes stars to be unbound, which in turn increase the apparent 
(empirically-derived) dynamical mass. When our TDG model is completely 
stripped of gas, its stars are completely unbound, and form an expanding 
and rotating envelope. In this situation, measurements of the dynamical 
mass may be substantially greater than the true mass by factors of 
$\sim$10. However, when the gas is not completely stripped and a bound 
stellar disk remains, the effect of unbound stars is much weaker. In this 
case, our dynamical masses are typically no more than double the true 
mass.

\section{Discussion}
\label{Discussion}

In this study, we have demonstrated that a lack of dark matter in tidal 
dwarf galaxies, coupled with high gas fractions, makes them very sensitive 
to ram pressure stripping. At radii where the gas is stripped from the 
disk, the stars are unbound. Stellar streams are often attributed to the actions of tides. Here we show that even ram pressure may play a role in shaping them. In fact, if the gas is completely stripped, 
we find that the TDG models are destroyed altogether.  We find it 
difficult to see how TDGs can survive the loss of their disk gas unless 
the relative densities of gas to stars within the disk is much lower than 
we have assumed.

Based on these results, TDGs that survive by maintaining their gas disks 
against ram pressure can perhaps be expected to have undergone particular 
evolutionary scenarios. They may have formed with small, relatively high 
surface brightness disks or have large masses. This provides them with 
greater disk self-gravity such that they are better able to hold onto 
their gas. Alternatively, they may form at large radii from their 
interacting progenitor galaxies, where the hot gaseous halos are low 
density. Their subsequent orbits may also be constrained so as to avoid 
plunging orbits past the progenitor galaxies, where combined high orbital 
velocity and high densities of hot gas could result in strong ram 
pressures, and therefore destruction. Plunging orbits near to the 
progenitors could also result in strong tidal forces, causing tidal 
stripping of stars and gas -- an additional mass-loss effect which we have 
not included in our models. Some TDGs may have only recently formed, so 
there has been insufficient time for the gas to be stripped, or prior to 
the first pericentre of their orbit past the progenitor galaxies. We note 
another possibility -- some TDGs may actually have been stripped of their 
gas, and been completely unbound. Our TDG models that suffered this fate 
do not show clear indications of being unbound, either visually, or in 
their stellar dynamics, beyond a steady decrease in surface density with 
increasing effective radius.

In fact, we suspect that a strong dynamical response of stars to gas 
removal by ram pressure is not limited to the TDGs themselves. The tidal 
streams of stars and gas that give birth to, and feed, newly formed TDGS 
should be even more sensitive to ram pressure. These streams are also 
expected to be gas rich, and are not dominated by dark matter. Given their 
low surface densities, and so weak self-gravity, they should be stripped 
of their gas for weaker ram pressures than the TDGs themselves. Even if 
ram pressure is unable to strip the gas from the TDGs, it may still play a 
role in altering a TDGs evolution by stripping tidal streams, therefore 
cutting off fresh supplies of gas and stars.

What might we expect if a tidal streams suffers ram pressure? In fact 
given the destructive impact on our TDG models, it is likely that the 
stars in an initially gas rich tidal tail would be be dissipated following 
ram pressure stripping. In this scenario, stellar-only streams would 
generally only be produced by the tidal dissolution of a galaxies, as ram 
pressure stripping may be too destructive to the stellar component of a 
gas rich tidal stream.

Given that tidal streams should be even more sensitive to ram pressure 
than TDGs, this raises the question: why are the gas and stars so well 
aligned in many tidal streams? Examples include the M81 group 
(\citealp{Yun1994}), NGC4676 (\citealp{Hibbard1996}), NGC2992/3 
(\citealp{Duc2000}), and NGC4038/39 (\citealp{Hibbard2001}). Although 
offsets can be found between the gas and stellar distribution in some 
tidal streams (e.g., Arp~299; \citealp{Hibbard1999}), these can be largely 
explained by differences in the initial distribution of the gas compared 
to the stars, and the dissipational nature of the gas compared to the 
dissipationless nature of the stars (\citealp{Mihos2001}). One possibility 
is that the streams can only exist where the ram pressure is weak -- at 
large radii where the hot gas density is low, and the velocity of the 
streams relative to the hot gas is low. The streams of cold gas may also 
temporarily be able to shield themselves by being enveloped in a their own 
hot gas that moves with the stream. This envelope must first be stripped, 
before cold stream can directly feel the ram pressure. A similar process 
is proposed to occur in disk galaxies within the galaxy cluster 
environment (\citealp{Bekki2009}). Alternatively, the hot gas halos may be 
flowing themselves, and gas streams may only form where they do not oppose 
the flow of the hot gas. An understanding of the dynamics of the hot 
gaseous halos of interacting galaxies may be vital for understanding the 
evolution of tidal streams and TDGs.

If we assume that TDGs and tidal streams are very sensitive to ram 
pressure, then we can use them as probes of their local environment. If, in 
general, they show no indication of ram pressure, perhaps the hot gaseous 
halos surrounding the TDGs are much lower density than we have been 
assuming. This would be very unexpected. For example, we choose the 
density of hot gas in our simulation to be typical of the measured density 
of the MW gaseous halo (\citealp{Bregman2007}; \citealp{Lehner2011}; 
\citealp{Gupta2012}). In fact, we might expect significantly higher hot 
gas densities in strongly interacting galaxies where compressive tidal 
forces and supernovae feedback may convert large quantities of cold disk 
gas into hot halo gas. Most TDGs form between two interacting gas rich 
disk galaxies (\citealp{Kaviraj2012}). If tidal streams and TDGs indicate 
that most external disk galaxies do not contain hot gas halos, then 
perhaps this indicates that the MW and/or the Local group is atypical.

A lack of hot halo gas in many late-type disk galaxies could have 
far-reaching consequences for disk galaxy evolution in general. 
Significant quantities of hot gas in a halo are a natural consequence of 
galaxy formation and evolution in $\Lambda$CDM (\citealp{Courty2010}), and 
a restocking of cold gas via cooling-out from a hot gas halo is required 
to maintain star formation rates, reproduce observed metallicity gradients 
(\citealp{Pilkington2012a}; \citealp{Gibson2013}), and produce realistic 
disk morphologies (\citealp{Hambleton2011}; \citealp{Stinson2012}). Our results may also have implications for the proposed TDG origin of the Local Group dwarf galaxies (\citealp{Metz2009}; \citealp{Kroupa2010}).

\section{Summary and Conclusions}

Tidal dwarf galaxies (TDGs) contain little or no dark matter. As such, 
they might be expected to be highly sensitive to their environment. In 
this study, we wish to understand the impact of ram pressure on TDGs. Ram 
pressure could arise from the motion of a TDG through the hot gaseous 
halos of its progenitor galaxies. Alternatively the ram pressure may arise 
if TDGs, which evade destruction by their progenitor galaxies, 
subsequently enter a group or galaxy cluster environment.

We submit TDG models, consisting of an exponential disk of gas and stars, 
to wind-tunnel ram pressure stripping tests. We conduct a parameter study, 
varying properties of the TDG models (such as mass, size, and gas 
fraction), and also varying the strength of the ram pressure.

We use a `toy' ram pressure model, and wind tunnel tests to study the 
effects of ram pressure on TDGs. Although this approach is highly 
idealised, we emphasise that the mechanism behind our key result -- the unbinding of stars -- 
is primarily the removal of the potential provided by the gas. Disk acceleration by ram pressure drag can further enhance stellar losses, but only if the loss of the gas potential first perturbs the stars. 
By definition ram pressure stripping involves the removal of the gas. 
Therefore this study highlights the sensitivity of tidal dwarf galaxies to 
ram pressure.

Our key results may be summarised as follows. 
\begin{enumerate}
\item The lack of a dark matter halo makes TDGs very sensitive to ram pressure.
\item At radii in the TDG disk where the gas is ram pressure stripped, the stars are unbound. This causes the gas {\it{and stellar}} disk to be truncated. If all of the gas is stripped, our TDG models are entirely destroyed.
\item Ram pressure causes a drag force on TDGs that accelerates them. Acceleration enhances star losses beyond that which occur from the loss of the gas mass alone. Acceleration can also cause unbound stars to lie in a low surface brightness stellar stream that is uncorrelated with the stripped gas stream.
\item For weak ram pressures, truncation of the stellar disk causes the surface density profile to have a reduced effective radius. For strong ram pressure that quickly sweeps out the gas, the stars are unbound and form an expanding envelope. In this case the effective radius steadily grows with time, while the surface density decreases. The Sersic index is only weakly affected.
\item The stellar dynamics of partially stripped TDG models provide dynamical masses that are within a factor of 2 of the real mass. However the stellar dynamics of TDG models that lose all their gas provide highly inflated dynamical masses -- up to $\sim$10 times the true mass. 
\end{enumerate}

TDGs are expected to contain little or no dark matter. Therefore they 
might be expected to be very sensitive to their environment. This includes 
external tides from other galaxies or, as this study demonstrates, ram 
pressure. Those TDGs that survive these multiple survival hurdles, might 
be expected to have very different properties from those with which they 
originally formed. We suggest that the role of ram pressure in simulations 
that form TDGs be considered carefully. The strong response of TDGs (and 
presumably the tidal tails from which they formed) to ram pressure make 
them sensitive probes of their local environment, and may provide 
constraints on the hot gas content of interacting galaxies.

\section*{Acknowledgements}
MF acknowledges support by FONDECYT grant 1095092 and FONDECYT grant 
1130521, RS acknowledges support by FONDECYT grant 3120135, and GC 
acknowledges support by FONDECYT grant 3130480. YKS acknowledges support 
by FONDECYT grant 3130470. BKG acknowledges the support of the UK Science 
\& Technology Facilities Council (ST/J001341/1), and the generous visitor 
support provided by FONDECYT and the Universidad de Concepcion.

\bibliography{bibfile}

\bsp

\label{lastpage}

\end{document}